**ABSTRACT:**

This article explores a critical gap in Mixed Reality (MR) technology: while advances have been made, MR still struggles to authentically replicate human embodiment and socio-motor interaction. For MR to enable truly meaningful social experiences, it needs to incorporate multi-modal data streams and multi-agent interaction capabilities. To address this challenge, we present a comprehensive glossary covering key topics such as Virtual Characters and Autonomisation, Responsible AI, Ethics by Design, and the Scientific Challenges of Social MR within Neuroscience, Embodiment, and Technology. Our aim is to drive the transformative evolution of MR technologies that prioritize human-centric innovation, fostering richer digital connections. We advocate for MR systems that enhance social interaction and collaboration between humans and virtual autonomous agents, ensuring inclusivity, ethical design and psychological safety in the process.

# Breaking the mould of Social Mixed Reality - State-of-the-Art and Glossary

**AUTHORS:**

Marta Bieńkiewicz *EuroMov Digital Health in Motion, Univ. Montpellier IMT Mines Ales, Montpellier, France*

Julia Ayache - *EuroMov Digital Health in Motion, Univ. Montpellier IMT Mines Ales, Montpellier, France*

Panayiotis Charalambous - *CYENS – Centre of Excellence, Cyprus*

Cristina Becchio - *University Medical Center Hamburg-Eppendorf, Hamburg, Germany*

Marco Corragio – *Scuola Superiore Meridionale, Naples, Italy*

Bertram Taetz - *Deutsches Forschungszentrum für Künstliche Intelligenz and International University of Applied Sciences, Germany*

Francesco De Lellis - *University of Napoli Federico II , Italy*

Antonio Grotta - *Scuola Superiore Meridionale, Naples, Italy*

Anna Server – *Vall d'Hebron Barcelona Hospital Campus, Spain*

Daniel Rammer - *Ars Electronica Futurelab, Linz, Austria*

Richard Kulpa – *Univ. Rennes, Inria, M2S, Rennes, France*

Franck Multon – *Univ. Rennes, Inria, CNRS, IRISA, Rennes, France*

Azucena Garcia-Palacios – *Dpt Psicologia Basica, Clinica y Psicobiologia. Universitat Jaume I, Spain*

Jessica Sutherland - *De Montfort University, Leicester, UK*

Kathleen Bryson - *De Montfort University, Leicester, UK*

Kathleen Richardson - *De Montfort University, Leicester, UK*

Stéphane Donikian – *Univ. Rennes, Inria, CNRS, IRISA, Rennes, France*

Didier Stricker – *DFKI - Deutsches Forschungszentrum für Künstliche Intelligenz and University of Kaiserslautern-Landau, Germany*

Benoît Bardy - *EuroMov Digital Health in Motion, Univ. Montpellier IMT Mines Ales, Montpellier, France*

## 1.  INTRODUCTION

We are living in an era marked by profound sociotechnical transformations, where societal challenges and rapid technological advancements are reshaping interactions in digital spaces toward increasingly multimodal forms of collaboration. Pressures such as the need to reduce carbon footprints, the impacts of global pandemics, and the rise of international and multilateral corporate structures are driving a shift from co-located collaboration to asynchronous, distributed platforms (e.g., Slack, Microsoft Teams,Discord). Simultaneously, technological progress—particularly in Artificial Intelligence (AI), propelled by machine learning and the widespread deployment of Large Language Models (LLMs) such as ChatGPT and Gemini—is accelerating. The convergence of AI, computer vision, digital twinning, distributed and edge computing, along with the expansion of mobile broadband networks, is significantly broadening the scope of what is technologically feasible. These developments enable higher-fidelity representations of human behavior within distributed virtual environments.

This convergence underpins the evolution of Mixed Reality (MR), also known as eXtended Reality (XR), which encompasses Virtual Reality (VR) — immersing individuals in fully virtual environments — and Augmented Reality (AR), which overlays virtual objects onto the real world (Milgram & Kishino, 1994). Despite its transformative potential, the MR revolution remains a work in progress rather than a fully realized phenomenon (Skarbez et al., 2023; Samala et al., 2023). One of its most pressing challenges is addressing the lack of embodiment and sociality that characterizes "traditional" physical environments.

Movement is fundamental to reality-based human interactions, as even speech relies on movement for its production. Social neuroscience consistently highlights the complex dynamics of multimodal signals embedded in physical and social presence (Bieńkiewicz et al., 2021). However, in virtual environments, movement representation typically lacks the richness inherent in the socio-motor interactions observed in the real world, which allow for social connectedness.

Currently, motion rendering techniques in MR are limited, typically tracking only one or two segments of the human body, such as handheld controller or headset. These methods fall short of providing the fidelity of a full-body motion capture system, restricting the depth and authenticity of virtual interactions. This challenge of translating "personalized" motor components, such as individual motion signatures (Słowiński et al., 2016), into MR underscores the complexity of achieving realistic human movement in virtual spaces. However, advancements in motion tracking techniques are poised to bridge this gap and will soon enable socially rich interactions, transcending the physical limitations of time and space.

A new era of Embodied Social Interaction in MR platforms is emerging, driven by research programs like Horizon Europe (e.g., Human-01-CNECT). This fast-approaching era of innovation is forecasted to revolutionize communication, allowing for novel modes of expression by transmitting key signals critical for efficient information exchange in amplified or symbolic forms. These advancements will not only enhance MR's usability beyond home gaming but also address key barriers to broader adoption, such as high equipment costs and specific usability needs in fields like education and healthcare. Furthermore, as MR technology evolves, so too must the ethical frameworks that guide its development. These initiatives are aligned with a human-centered design principles, focusing on empowering individuals through technology that is decentralized, inclusive, and positively impacts society. Therefore, these frameworks will need to incorporate guidelines for designing human-computer interactions with embodied AI agents, emphasizing usability, acceptability, and responsible integration into diverse societal contexts.

This manuscript highlights the recent advancements, current challenges and ethical and societal implications of rendering and generating artificial movement in virtual environments. First, we provide a brief state of the art of motion rendering within MR technologies. Then we introduce a curated glossary of 23 key-terms, mapping the technical, scientific and ethical challenges of breaking the barrier of socially relevant artificial movement generation in MR. Finally, we discuss the implications of these advancements for the development of MR in key societal domains.

## 2.    STATE OF THE ART IN SOCIAL MR

This section examines recent advancements in the collective technological field of MR to identify potential pathways for realizing its transformative potential in enabling embodied interactions between humans and AI on digital platforms. We begin by providing an overview of MR platforms and computer-mediated collective work platforms (groupware), highlighting their purposes and capabilities in mediating human-to-human interactions. We then explore how these platforms could evolve to support interactions with embodied, autonomous virtual agents (L3) in the future.

2.1 MR-embodied path to sociality

In recent years, the Metaverse has garnered significant attention, driven by Facebook's (now Meta) investment and its promise of a permanent, immersive digital reality unprecedented in scope (Mystakidis, 2022). A defining feature of the Metaverse is its accessibility, offering a parallel reality free from the physical world's spatiotemporal constraints. Within this interconnected and interoperable space, users interact through personalized avatars, enabling freedom of social identity and equal participation (Suk and Laine, 2023).

The Metaverse also introduced alternative power structures, empowering users who possess immersive technology and technical skills to create, exchange services, and generate income on equal footing. Despite its ties to dominant tech industries, it was envisioned as a counterbalance to traditional revenue and power consolidation (Bibri et al., 2022). However, its current usage remains limited, driving revenue from gaming and e-commerce, without eliciting mass adoption yet.

A critical gap in MR, including the Metaverse, is the absence of socio-motor components essential for embodied, socially rich interactions between humans and autonomous virtual characters. While groupware has facilitated joint tasks in digital spaces, its capabilities remain limited to explicit, often symbolic gestures, such as those conveyed through emojis or basic animations (Ens et al., 2019). Those movement representations, as we referred to as "para-movement," lacks the rich social information, inherent in natural human interactions, which is difficult to reconstruct through video streaming or simplistic digital proxies. Luo et al. (2022) emphasize the importance of designing platforms with considerations such as spatiotemporal constraints, symmetry in exchanges, interaction scenarios, and attentional focus between agents and artificiality.

Since Johansson's (1973) pioneering work, it has been clear that biological motion is perceived differently from mechanical motion (Blake & Shiffar, 2007; Chaminade et al., 2007). Research suggests this distinction is shaped by evolutionary adaptations for detecting biological motion critical to survival and reproduction (Bryson, 2017), as well as anthropomorphism, which influences how we attribute intentions and agency to moving entities (Mar et al., 2007; Pavlova, 2012). Nevertheless, debates persist about whether these perceptions are primarily driven by top-down processes (e.g., social expectations) or bottom-up cues (e.g., motion subtleties) (Blake & Shiffar, 2007).

Evidence shows that humans can differentiate between movement patterns generated by avatars (controlled by humans) and autonomous virtual characters, with a preference for the former (Fox, 2015; Kelso et al., 2009). Advances in AI-driven motion generation and photorealistic datasets could help autonomous characters appear more naturalistic and human-like (Blascovich et al., 2002; Gratch et al., 2007). Enhancements, such as incorporating intention layers into autonomous character behaviors, have demonstrated that people can distinguish between declarative, informative, and imperative gestures during computer-mediated interactions (Raghavan, 2023). Adding these elements in real-time MR settings could create more embodied and meaningful interactions (Riva et al., 2019).

However, challenges remain. Improved graphical fidelity and human-like motion risk triggering the "uncanny valley" effect, where hyper-realistic but imperfect representations elicit discomfort (Mori, 1970). Balancing naturalism with user comfort will be crucial for the successful integration of autonomous virtual characters into MR environments.

## 2.2 Movement will socialize AI

At present, human interaction with AI engines occurs primarily through disembodied formats, such as text or voice prompts, lacking the multimodal signals necessary for richer, more intuitive communication. Advances in AI are rapidly paving the way for new multimodal technologies, enabling the real-time embodiment of multiple agents (e.g., Mok, 2023). To forge this next step forward, autonomous virtual characters must be trained on socio-motor features, allowing them to engage in meaningful, human-like interactions. Insights from social neuroscience, which explores the multimodal dynamics of human social behavior (Dumas, 2011; Pan & Hamilton, 2018; Schilbach et al., 2013), provide a foundation for transferring these interaction capacities to embodied AI agents. MR environments offer a unique platform for hosting such interactions, leveraging AI trained on multimodal data to create dynamic, intuitive, and socially rich exchanges.

Autonomous virtual characters are driven by different traditional AI techniques, but unlike avatars are not steered and are not playable by humans (Fox et al., 2015, Hennig-Thurau, 2022). These autonomous agents hold significant promise for conversational interactions with humans, a capability anticipated as AI technology progresses (Gal & Grosz, 2022).

The deployment of such agents offers considerable economic incentives, particularly in industries seeking cost-effective labor alternatives. Autonomous agents do not require compensation or contracts and are unaffected by human limitations, such as frustration—assuming they remain non-sentient, a topic of ongoing debate within the AI community (Mitchell & Krakauer, 2023). Additionally, these agents can be interchangeable and less susceptible to social judgments, potentially allowing users to deviate from typical social norms, such as maintaining personal space (Bailenson et al., 2003).

However, evidence suggests that humans often apply similar social norms when interacting with computers as they do with other people, reflecting ingrained social behaviors (Nass et al., 1994). This underscores the complexity of integrating autonomous virtual agents into human-centric environments while maintaining natural and meaningful interactions.

Assuming future artificial agents are endowed with autonomy or even agency, as envisioned by Gaggioli et al. (2016), this opens possibilities for symbiotic human-machine experiences. While the inevitability of developing human-like agency by AI remains speculative, humans already interact with technology as if it possesses agency—whether through robots or assistive devices (Richardson, 2015; Bryson, 2017). This paradigm of confluence suggests a shift beyond traditional technological impacts, emphasizing the co-evolution and resonance between humans and AI systems (Lee et al., 2022). However, substantial research is still required to understand these dynamics fully.

The acceptability of embodied, autonomous virtual characters in MR environments remains uncertain. Key questions arise: How will repeated interactions affect human emotional and psychological states? Will such interactions carry the same social weight as human-to-human exchanges? How will movements generated by AI compare to those rendered by humans in digital spaces? These uncertainties underscore ethical considerations surrounding AI-human interaction in MR.

## 3. CHALLENGES IN SOCIAL MR

To address these important questions and to foster meaningful technological progress, the development of transformative technologies must be participatory (Oudhof, 2023) and driven by ethical design (Tasioulas, 2022), prioritizing the dignity and well-being of humans and other animated entities (Rotenberg, 2021). Therefore, establishing a shared language—avoiding overlapping terminologies— is vital for advancing the field (Tiedrich et al., 2023). Hence, we introduce a Glossary of 23 essential terms currently missing from MR systems that are crucial for their development. Those terms were selected for supporting key design considerations for mitigating potential risks and harms in addressing the technical, scientific and ethical challenges of social MR (see Figure 1 for an overview).

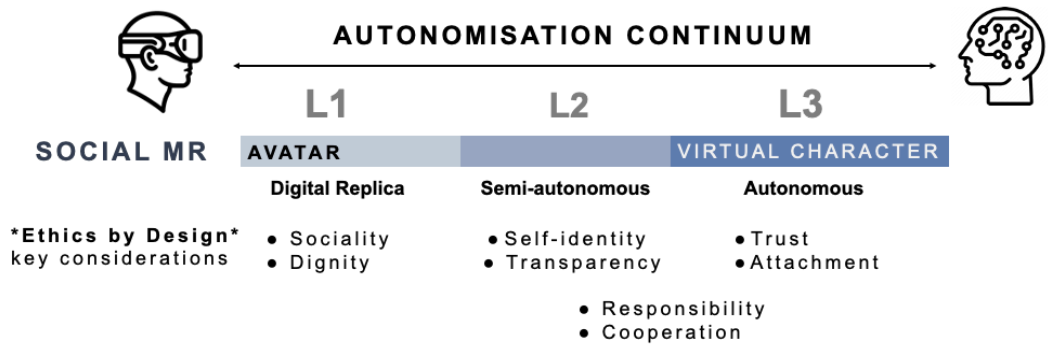

**Figure 1:** Graphical illustration of the autonomisation continuum and the ethical considerations that need to be addressed.

3.1 Technical challenges

The first and second sections of the Glossary introduces key-terms associated with the autonomisation continuum and technical challenges such as handling high volumes of multi-modal information for capturing/rendering scene and motion, reducing latency leading to cybersickness and managing various level of autonomy.

Multi-modal VR-systems are at present cumbersome and struggle with synchronizing multiple data streams in real-time. For instance, Ha et al. (2022) showcased a potential platform interface with three people accessing the system, but live streaming of motion capture data from all agents was not possible, limited to one active agent at once.

Recent technological advances in rendering algorithms and real-time global illumination in particular. Deep learning models such as super-resolution (e.g., Nvidia's DLSS), low-power and efficient mobile VR platforms (e.g., Meta Quest, Pico) and the high computation capabilities of Graphics Processing Units (GPUs) allow for the efficient rendering of very high-fidelity images at high frame rates on commercially available VR/XR platforms (see Scene neural rendering term in Glossary). Intermediate platforms designed to reduce latency, as discussed by Villagran-Vizcarra (2023), also show potential in addressing these challenges, paving the way for more seamless and immersive MR experiences.

The wireless sensors currently used to capture human movement in real-time for MR platforms face significant precision challenges. High-fidelity movement capture and rendering require substantial processing power, which introduces delays during real-time capture (Karuzaki et al., 2021). Traditional methods for capturing 3D body data often demand complex and costly technical setups, while more affordable alternatives typically fail to deliver the required data quality (Milosevic et al., 2020). AI-driven solutions have been identified as promising avenues to address these limitations, particularly by reducing latency and bandwidth demands for streaming and rendering real-time emotional expressions (see ). These advancements could significantly improve the realism and social dynamics of MR interactions.

Several challenges specific to Head-Mounted Displays (HMDs) persist, including cybersickness, latency, and low motion precision. HMDs were initially theorized to mitigate issues such as eye haze (Bailenson, 2021) and provide greater freedom of motion via solutions such as ego-centric visual-inertial tracking (see Ego-centric visual-inertial tracking in Glossary). However, cybersickness, though not directly diminishing immersion (Page et al., 2023), remains a significant concern. Cybersickness manifests through symptoms ranging from nausea and oculomotor strain to disorientation, with individual susceptibility shaped by multiple factors. These include posture (Merhi et al., 2007), sensitivity to motion sickness (Kourtesis et al., 2024), degree of control (Chen et al., 2012), prior experience (Weech et al., 2020), and demographic variables such as sex (Munafo et al., 2017) and age (Ars & Cerney, 2005), although the latter two remain subjects of ongoing research (Bailey et al., 2022; Li et al., 2023). Additionally, mismatches between the physical and digital body during VR exposure can lead to re-adaptation discomfort upon returning to reality (Maloney et al., 2019).

Finally, the development of artificial agents driven by AI introduces a automation continuum whereby AI-driven agents could facilitate nuanced interactions between humans and artificial Movement representation term in Glossary, including embodied autonomous or semi-autonomous virtual characters. Semi-autonomous virtual characters (L2), driven by humans but enhanced with AI-based transformations, and virtual autonomous characters (L3) with complete embodiment might redefine virtual interactions (see Automation continuum in Glossary).

In summary, addressing gaps in kinematic streaming, motion capture and rendering, and cybersickness mitigation is critical for enhancing realistic socio-motor interactions in MR environments. Beyond these technical challenges, unresolved scientific questions also demand attention.

3.2 Scientific challenges

The third section of the Glossary introduces key-terms associated with the scientific understanding of the role of multi-modal signalling in collaborative multi-agent scenarios. There is current lack of understanding of the impact of manipulating sociomotor components of social interactions, despite their importance in conveying social information such as intentions and emotions through sensorimotor propagation and synchronisation (Riva et al., 2011; Bieńkiewicz et al., 2021).

Sociality in MR needs to include movement for (i) fostering a sense of embodiment, (ii) social presence and (ii) for exchanging social information via propagation and/or synchronisation, with humans interacting with each other as avatars (L1) or with (semi-) autonomous virtual agents (L2/L3) (Lombardi et al., 2021; Oh et al. 2018; Bailenson, 2021).

Sense of embodiment can be manipulated through photorealistic representations that enhance immersion, body ownership, and social presence, as well as encode finer-grained gestures that enrich the social dimensions of movement (Latoschik et al., 2017; Weidner et al., 2023, Kilteni et al., 2021). True embodiment, safeguarding dignity, requires higher customization, including features like clothing, sex, ethnicity, and body shape, to enhance body ownership and achieve greater visual and behavioral fidelity. Synchrony between the digital and physical body is essential for full immersion, as the perception-action loop must align with the consequences of one's movements in the MR environment (see Synchronisation in Glossary). Additionally, avatar's (L1) anthropomorphic features must match the user's real body to improve task performance in virtual settings (Weidner et al., 2023). However, full-body representation is not always necessary, particularly in goal-oriented scenarios, where partial representations may suffice (Suk and Laine, 2023). Hence, a trade-off exists between photorealism and the ability to focus on kinematic movement components, as simplified representations are often better suited for capturing movement features (Chaminade et al., 2007; Zibrek et al., 2019). In contrast, mirror-like realistic representations could reduce embodiment and attentional focus, suggesting that overly realistic avatars may sometimes be distracting (Döllinger et al., 2023).

Socio-motor embodiment has been identified as critical to user experiences in digital spaces. For instance, social presence (see Social Presence in Glossary, and survey by Skarbez et al., 2017) increases with the use of HMDs and is further enhanced when autonomous virtual characters exhibit human-like behaviors, such as gaze imitating visual attention to their environment (Kim et al., 2019; Voinea, 2022). Non-behavioral environmental factors also influence interaction plausibility. Gentle airflow or the seamless blending of virtual and real elements has been shown to enhance user reception and interaction quality (Kim et al., 2019). As autonomous virtual characters (L3) grow increasingly human-like across all levels of automation continuum, their interactions with humans become more plausible and socially meaningful (see Social Connectedness in Glossary). However, as previously discussed, capturing and rendering emotions and intentions in MR settings remain challenging, due poverty of the socio-motor information encoded in the displays. Some approaches explore synthetic experiences across multiple modalities—audio-visual, haptic, olfactory, and even neural activity recordings—but their integration remains underdeveloped (Rakkolainen et al., 2021) and disconnected from social neuroscience.

The representation of human-like bodies, whether as avatars (L1) or embodied autonomous virtual characters (L3), introduces complex issues in MR environments across the autonomisation continuum (Ayache et al., 2023). Current MR avatar solutions, as noted by Karuzaki et al. (2021), are largely predesigned or character-based, limiting the potential for fully embodied representations. This can be achieved by social MR technologies that foster amplification of sensorimotor primitives to encode social information in such a way that it facilitates the readout and encoding of social information by human interactants (Becchio et al., 2024) - (see Sensorimotor Primitives and Sensorimotor propagation and Amplification in Glossary). Motor primitives are a fundamental, modular and reusable patters of motor activity (Giszter, 2015) and allow our brain to encode actions and produce diverse motor behaviours, without controlling every single muscle for force and speed. The use of sensorimotor primitives is associated with movement characteristics and movement prediction (and was used successfully in robotics - see Morrow & Khosla, 1995), rather than emotion recognition, set as 'red lines' by many AI regulations (i.e., recital 18 EU AI Act).

These capabilities raise ethical concerns, particularly regarding responsibility for unregulated behavior in MR spaces. For instance, Ramirez (2023) highlighted the inevitability of sexual harassment in MR settings, as virtual representations can replicate harmful real-world behaviours. A recent case involving virtual harassment and sexual assault in the Metaverse underscores the need for accountability and mitigation strategies (Sales, 2024). The immersive and shared nature of MR spaces—spanning private, corporate, and public settings—necessitates the development of principles addressing safety, privacy, autonomy, and dignity. Without such measures, the potential for harm remains significant. The following section will explore the ethical and societal impacts of MR technologies in greater detail.

3.3 Ethical challenges

The fourth section of the Glossary introduces key-terms associated with ethical challenges in MR environments. These challenges are numerous and have been highlighted in previous studies (Maloney et al., 2019; Rueda & Lara, 2020; Tasioulas, 2022). However, these issues bear reiteration given the potential of MR to incorporate sociality as dimension and mediate future social interactions across the spectrum of automatisation.

Digital interactions fostering in MR environments must consider principles of social influence established by foundational works like those of Allport (1985) and Cialdini (2005). Human behaviour changes in the presence—real or perceived—of others, where the "other" does not need to be physically visible; an imagined, symbolic, or implied presence can similarly steer behaviour. This presence fosters conformity, compliance, and adherence to social norms. In groups, such dynamics often amplify attitudes and normative behaviours, which can lead to both positive and negative outcomes. This subsection focuses on three predominant concerns: (i) Breach of Transparency and Trust, (ii) Excessive Attachment through stickiness, and (iii) Data Protection.

Breach of transparency and trust lies at the core of MR experiences, primarily driven by the creation of a sense of presence and illusion leading to deception. According to Slater (2009), immersion in MR enables the sense of presence (the illusion of being there) and plausibility, both tied to the concept of Response-As-If-Real (RAIR). These elements are often induced through body illusions, where users experience a sense of ownership over a virtual body. Achieving this requires a precise alignment between participants' anthropomorphic data (the "body matrix"; Moseley et al., 2012) and the coherence of the virtual environment linked to user expectations (Skarbez et al., 2021). While addressing the technical and scientific challenges associated with these elements seems feasible, the ethical implications are far more complex (see Trust and Transparency in Glossary). Addressing the ethical dimensions of deception in MR is particularly challenging because the immersive nature of these experiences blurs the boundaries between real and virtual, heightening both their potential and their risks to human dignity (see Dignity in Glossary).

The design of "successful" immersive experiences often increases the risk of creating overly "sticky" products ('over resonant') that exploit human psychology, such as fostering excessive engagement or reliance on digital content (Lomas et al., 2022; Leiser, 2024) (see Attachment in Glossary). This tendency can have far-reaching societal consequences, both immediate and long-term. For example, overreliance on immersive technology could lead to workforce disruptions (Curtis et al., 2023) and potentially adverse cultural or behavioral shifts without mechanisms to mitigate these effects. Given the unpredictability of societal impacts from emerging technologies, a proactive approach to design is necessary—one that prioritizes social responsibility, safeguarding from overreliance on technology (attachment) or negative outcomes (Ayache et al., 2023).

Avatar representation (L1) also influences user behaviour, both positively and negatively. For example, the "Proteus effect" describes how avatar characteristics affect attitudes and behaviors. Virtual representations, such as sexualized female avatars, can promote self-objectification and behavioural changes in women users (Fox et al., 2013; Cikara et al., 2011). Hyper-realistic avatars can evoke prosocial behaviours in the real world (Ahn et al., 2013) but also identification to an idealized self leading to addictive behaviours as a result of affective attachment to their idealized virtual avatar (Szolin et al. 2023). High levels of presence in MR experiences may lead to embodied consequences beyond the virtual environment (Riva et al., 2019) - (see Self-Identity and Sociality in Glossary).

Currently, many MR applications remain experimental and are primarily used as research tools, given the limitations in interactional embodiment. However, as MR systems become more integrated into daily life, the rapid pace of technological innovation is likely to outstrip the development of regulatory frameworks capable of safeguarding users (Leiser, 2024). This emphasizes the need for **ethics by design**, where ethical considerations are embedded throughout the development process rather than being reactive measures, to promote beneficial design. This approach enables anticipatory identification of potential harms and establishes safeguards as part of the system's functional requirements (see Responsibility and Cooperation in Glossary).

Social MR platforms, like the Metaverse, highlight this challenge. Presently, these systems lack robust mechanisms for user accountability or compliance with privacy regulations such as GDPR (Cheong, 2022). Incorporating ethical guardrails into these platforms before deployment can help prevent misuse and protect user dignity and well-being. For instance, blockchain technology can safeguard user identity and privacy, ensuring a more secure and trustworthy interaction environment (French et al., 2021). Additionally, interoperable and hardware-agnostic platforms supported by shared 5G access may democratize participation, allowing for a more equitable and inclusive user experience (Mystakidis, 2022).

In summary, ethical considerations are integral to the development of MR platforms (Slater, 2020). Deception, misinformation, and impersonation—already challenges in real-world interactions—become amplified in digital environments, posing risks of antisocial behaviour and harm (Grinbaum & Adomaitis, 2022). Therefore, protective measures must be established, ensuring these spaces remain safe and empowering for all users (Kaddoura & Al Husseiny, 2023). This becomes especially crucial for semi-autonomous virtual characters (L2), which introduce questions about one's self-identity and transparency.

Anticipatory governance such as the EU Act should prioritise these ethical concerns before market deployment (Yang, 2023), recognizing that MR technology has potential to reshape societal paradigms and bridge gaps between science, technology, and human values (Bibri et al., 2022). As a research community, our responsibility is to foresee and mitigate potential risks, particularly linked to AI-optimised designs (Leiser, 2024). Technology is not value-neutral; it reflects and shapes societal values, often tilting toward industrial profit motives such as advertising and subscriptions (Kudina & Verbeek, 2019; Varoufakis, 2023). The challenge is to ensure that human values take precedence, guiding the development of immersive platforms toward a more ethical and inclusive future. In the next section, we discuss the implications of these advancements for the development of MR in key societal domains.

## 4. <u>KEY AREAS FOR SOCIETAL APPLICATIONS</u>

In this section, we discuss the current challenges in deploying MR technologies across key societal areas such as healthcare, sports, arts and education, highlighting specific opportunities for benefits brought by technology development towards embodied social MR.

### 4.1 Mixed Reality in Healthcare

The deployment of VR in healthcare has largely been confined to research settings, with limited real-world implementation in healthcare services. This is primarily due to technical constraints of MR systems and challenges in drawing conclusive evidence about their efficacy, given rapid technological advancements and difficulties in assessing long-term impacts (Sokołowska, 2023). Additionally, adapting MR technology to healthcare settings poses unique obstacles, such as sensors for capturing body movements performing poorly with individuals with disabilities (Aufheimer et al., 2023).

Despite these challenges, MR environments offer significant potential for clinical applications. They enable the replication of naturalistic scenarios while providing precise control over variables, which is particularly promising for neuropsychological assessment and neurorehabilitation. For example, MR has shown potential in aiding patients with traumatic brain injuries or neurodegenerative disorders by enabling assessments that blend functional testing with daily life behavior (Parsons et al., 2020). However, the cognitive and perceptual thresholds required for neurological patients to effectively engage with MR applications remain unclear (Calabrò et al., 2022).

While VR approaches already contribute meaningfully to neurorehabilitation, advancements such as AI-driven scene neural rendering could enhance these systems further. For instance, immersive environments could allow a single therapist to oversee multiple patients simultaneously, tailoring interactions in real time without compromising care quality. MR solutions featuring avatar therapists are projected to address diverse therapeutic needs, including motor planning rehabilitation, chronic pain alleviation, and symptom management for psychiatric disorders (Sokołowska, 2023). Early-stage research into MR physical rehabilitation has demonstrated significant improvements in user outcomes, with fidelity and motivational factors identified as critical for success (Howard & Davis, 2022).

MR also shows promise in addressing chronic low back pain, a leading cause of disability worldwide. Studies indicate VR interventions can significantly reduce pain intensity and kinesiophobia, offering a viable option for patients with limited therapeutic alternatives (Brea-Gómez et al., 2021; Hayden et al., 2021). Moreover, remote MR rehabilitation can enhance accessibility by reducing the need for travel to clinical centers. Shared MR experiences further enable group therapy sessions enriched with sensor-based feedback, promoting holistic treatment approaches that foster social connectedness and positive emotions (Bieńkiewicz et al., 2021; Smykovskyi et al., 2022).

However, several challenges remain unresolved around technical aspects of movement representation in MR. For example, patients tend to move more slowly in MR spaces compared to real-world environments, potentially limiting the transferability of MR-based training to meaningful clinical outcomes (Arlati et al., 2022). Nonetheless, specific applications, such as virtual imagery therapy for paralysis, are unaffected by this limitation. Rendering desired expressions for patients with facial paralysis has been shown to stimulate motor representations, aiding recovery (Bernd et al., 2018). Similarly, MR can help patients adapt to anticipated treatment trajectories, such as through exposure therapy in cancer care (Sansoni et al., 2022), and VR-based exercise therapy has demonstrated effectiveness in chronic pain management (Bilika et al., 2023).

A key limitation of current MR therapeutic applications is the insufficient design of interactions between therapists and patients, limiting experience of social presence. Unlike in game design, where users operate autonomously, therapists in MR settings must guide participants while encouraging self-determination (Aufheimer et al., 2023). Effective therapeutic interventions require a patient-centered approach that accounts for emotional vulnerabilities and emphasizes constructive feedback. This contrasts with gaming environments, which often emphasize performance comparisons and can lack the supportive feedback essential for rehabilitation.

Future developments in embodied social MR, drawing on neuroscience of sensorimotor primitives (Becchio et al., 2024), hold great promise for overcoming these limitations. Integrating autonomous virtual characters and human avatars in embodied, socially rich therapeutic contexts could enable more interactive and personalized care. Social MR environments may also facilitate scenarios for collaborative exercises (through means of sensorimotor primitives amplification methods), enhancing motivation through autonomy while preserving the essential human-led nature of therapy. By addressing current challenges, MR technologies can evolve into a transformative tool for healthcare.

4.2 Mixed Reality in Sport

MR technologies hold significant potential for sports and dance applications, offering innovative avenues for training, performance enhancement, and engagement. VR enables controlled, repeatable environments for skill refinement, coaching emulation (Cojocaru et al., 2022; Mystakidis et al., 2023), and post-impact health monitoring (Craig et al., 2022). In dance, VR fosters fun, fitness, and social interaction while reducing stress and physical constraints, enhancing motivation and creativity (Sarupuri et al., 2023). For sports, the focus shifts to motor skill training and the critical transfer of virtual skills to real-world performance (Bideau et al., 2010).

Despite its promise, MR deployment in sports faces significant technological challenges. Issues such as low frame rates, refresh rates, and the absence of haptic feedback persist (Le Noury et al., 2023). Latency between user actions and their visual movement representation disrupts the perception-action loop, which is critical for motor performance, especially in high-speed activities (Hoyet et al., 2019; Morice et al., 2008). Even imperceptible delays can impair precision tasks like basketball free throws or long-distance aiming (Covaci et al., 2015). VR environments can also alter spatial perception, leading to locomotor adaptations that undermine training effectiveness (Pontonnier et al., 2014). Furthermore, rendering quality (Vignais et al., 2009) and interaction devices (Berton et al., 2019) influence biofidelity—the accuracy of replicating real-world dynamics—and skill transfer, necessitating careful evaluation of these systems.

Yet, VR has proven effective in enhancing specific sports skills, such as tracking multiple players in soccer (Vu et al., 2022), improving anticipation in handball goalkeeping (Vignais et al., 2009), and supporting opponent interception in rugby (Bideau et al., 2010). Fast-paced advancements integrate traditional kinematic metrics with user acceptance assessments (Mascret et al., 2022), while VR opens new possibilities like augmenting visual feedback to highlight key perceptual cues (Limballe et al., 2022) and automating athlete adaptations for personalized training (Gray, 2017).

Similarly, AR has shown effectiveness in real-world training, enhancing sport climbing through instructor-free repetitions (Heo and Kim, 2021) and replicating realistic shooting conditions by preserving recoil forces (Lucero-Urresta et al., 2021). AR applications extend to fan engagement, revenue diversification for sports clubs, and professional training (Sawan et al., 2020). Integrated with AI, AR offers adaptable and flexible tools for motor research and sports development (Solas-Martínez et al., 2023). However, challenges remain, such as ensuring real-time colocation of virtual and physical objects on accessible, affordable devices, and vigilance regarding privacy and data protection (Mehra et al, 2023).

The evolution toward embodied MR promises significant advancements by integrating real-time virtual interactions with physical dynamics (such as sensorimotor primitives), fostering both physical and psychological engagement. Realistic virtual agents, improved haptic feedback, and latency reductions (scene neural rendering techniques and ego-centric visual-inertial tracking**)** can enhance ecological validity, bridging the gap between virtual practice and real-world performance. By addressing these constraints, embodied MR can offer unprecedented opportunities for skill development, sustained motivation, and holistic athlete training.

4.3 MR in Arts

The application of MR in the arts is multifaceted, spanning from mobile AR applications on smartphones to installations using HMDs and shared experiences via large-scale projections where no individual devices are required. While single-user MR tools dominate artistic applications, a growing need exists for collaborative and multi-user experiences. For instance, *Self-Compass* (Goepel et al., 2023) demonstrates how MR can augment physical structures—merging physical and virtual realities to encourage visitors' exploration of place and experience. However, multi-user artistic projects remain limited, with examples like *MultiBrush* (Rendever, 2021) highlighting their potential.

MR is already well-established in the field of digital performance art, offering varying levels of participation for performers, audience members, and even virtual autonomous agents (Grasset, 2008). This flexibility enables a spectrum of engagement—from anonymous contributions via mobile devices to active participation on stage—creating novel blends of physical and virtual presence. The study Weijdom (2022) provides a conceptual framework for integrating MR into theatrical contexts, addressing scenographic design, audience interaction, and the role of technology in shaping performance aesthetics. This approach emphasizes the value of iterative, interdisciplinary collaboration among artists, technologists, designers, and researchers. These collaborative processes prioritize active audience engagement and dynamic creative exchange, enabling artistic disciplines to co-evolve with emerging technologies.

Audience participation has shown promising effects on engagement and creativity. Lindinger et al. (2013) examined participatory dance performances, where dancers interacted with real-time audience-generated text projected on a large screen. Findings suggest that principles such as offering users free choice to engage enhance creative expression and promote *Social Flow*—a collective sense of immersion and collaboration.

The concept of *Collaborative Aesthetics*, evolving from *Cooperative Aesthetics* (Funk, 2016), captures the unique potential of MR to facilitate shared artistic creation. Unlike cooperation, which focuses on achieving a common goal without individual alignment, *Collaborative Aesthetics* emphasizes co-creative processes where individuals contribute ideas and perspectives to shape a collective outcome. MR environments enable dynamic interaction, empowering groups to generate shared narratives or audiovisual art pieces that reflect both individual and collective creativity.

By bridging artistic vision, technology, and embodied experiences, MR offers transformative possibilities for collaborative art, fostering innovative, immersive, and participatory environments that redefine artistic creation and audience engagement.

4.4 Mixed Reality in Education

MR use for Education, despite intense research, large-scale deployment remains limited (Zabuli et al., 2023). Key barriers include high system costs, limited interoperability, and a lack of realism and multimodality compared to in-person education (Gonzalez-Moreno et al., 2023). While remote access through autonomous agents shows potential, challenges persist regarding user frustration and inconclusive evidence on the efficacy of autonomous characters versus avatars for educational purposes (Mystakidis and Lympouridis, 2023).

On one side, MR environments offer unique opportunities for sensory imagery and guided instruction during manual tasks. This is particularly valuable for craft education, where the combination of external material interactions and internal expectations forms the foundation of expert knowledge and semantic representations (Hauser et al., 2022; Coley et al., 2019). Attention to environmental somatosensory stimuli is critical for cultural heritage preservation and skill transfer in craft education. Here, MR facilitates not only the observation of movements but also the simulation of complex actions, making it an invaluable tool for skill development.

One of MR's most promising avenues lies in experiential learning—"learning by doing"—which can facilitate reorganization of episodic memory, an outcome often unattainable in conventional classroom settings. However, current MR systems face critical shortcomings in fostering emotional engagement and providing a full spectrum of multimodal embodiment, including vision, sound, spatial awareness, and smell. Emotional interest is fundamental for effective learning, yet MR has yet to fully replicate this immersive, multi-sensory experience (Mystakidis and Lympouridis, 2023).

Embodied MR technology, with possibility to amplify or attenuate sensorimotor primitives, holds immense potential to revolutionize education by bridging gaps in emotional engagement, multimodal embodiment, and accessibility, fostering social presence and connectedness. By integrating advanced sensory stimuli and reducing technological barriers, MR can enable richer, more interactive learning experiences. Nonetheless, challenges related to cost, privacy, and system interoperability must be addressed for widespread adoption. Continued research into the efficacy of embodied social MR will be critical to unlocking MR's full promise for education.

4.4 Digital Twins and Mixed Reality

Digital Twins (DTs) refer to virtual replicas or digital representations of physical objects, systems, or processes. These digital models can simulate, predict, and optimize the behavior and performance of their real-world counterparts in real time. To create a DT, data from sensors, devices, and other sources are collected and integrated into a 2D or 3D virtual representation, essentially creating a digital copy of the physical object (Wu et al., 2021; Barricelli et al., 2019). Initially, DTs were experimental technologies aimed at replicating the elements, functions, operations, and dynamics of physical systems in the digital realm. However, the supporting technologies were not advanced enough to handle complex systems or systems-of-systems.

Recent advancements in Artificial Intelligence (AI), Machine Learning, Mixed Reality (MR), sensing, security, cloud storage, transfer learning, data visualization, and ultra-reliable low-latency communications have made it possible to implement DTs and expand their applications across various industries. Once limited to isolated processes, DTs now have the capacity to replicate the processes, elements, dynamics, firmware, connections, and operations of entire physical systems. These DTs can then be used to monitor, control, and optimize performance while identifying potential problems and opportunities for improvement.

Various industries, including manufacturing, healthcare, aerospace, cultural heritage, and transportation, are increasingly adopting DTs to improve efficiency, reduce costs, and enhance safety and reliability. Smart city initiatives find DTs particularly valuable for infrastructure optimization, resource management, and public service improvements, ultimately enhancing the quality of life for citizens. Urbanism, in particular, represents a significant application of DTs, as they enable immersive exploration of city landscapes. These virtual replicas reflect cultural dynamics and power structures within societies, providing new tools for understanding and managing urban environments (Graham et al., 2022).

However, the deployment of DTs in urban spaces also highlights critical governance challenges. Privacy and data protection emerge as paramount concerns, as real-time data collection and replication involve sensitive information about individuals and communities. Addressing these issues will be essential to ensure ethical and secure deployment of MR and DTs, particularly in educational or public-sector applications (Allam et al., 2022).

The integration of virtual humans, with socially rich motor representations, both digital twins of real individuals and autonomous virtual characters—presents a significant opportunity within the MR space. This development could enable novel forms of interaction and collaboration, transforming MR environments into immersive, social, and highly interactive spaces while unlocking innovative ways to engage with DTs (Numan et al., 2023).

## 5. <u>CONCLUSION</u>

The introduction of a curated Glossary of key terms aims to provide a conceptual foundation for researchers, designers, and policymakers navigating this growing field. By promoting shared language and interdisciplinary alignment, we seek to bridge gaps between technological capabilities and societal needs. Ultimately, the future of MR lies in its capacity to support embodied, socially connected experiences that respect user dignity, privacy, and agency. Continued collaboration across scientific, artistic, and technological disciplines—as well as anticipatory ethical frameworks—will be essential to ensure MR technologies evolve inclusively and responsibly. As MR becomes more deeply embedded in daily life, its development must be guided not only by what is technologically possible but foremost by commitment to human wellbeing.

### <u>Acknowledgements:</u>

Co-funded/Funded by the European Union under Horizon Europe, grant number 101092889, project SHARESPACE. Views and opinions expressed are however those of the authors only and do not necessarily reflect those of the European Union. Neither the European Union nor the granting authority can be held responsible for them.

We want to thank all the partners of the SHARESPACE consortium for their collaborative work on the grant proposal (with initial Glossary terms) and contribution to D1.1 The SHARESPACE Unified and Operational Framework. This paper stemmed from this joint effort.

We want to thank Dr. Kathleen Bryson for editing and proofreading the manuscript, and Emmanuel Herbert for the insightful comments on the manuscript.

## **GLOSSARY KEY TERMS**

| Theme 1 - Autonomisation Continuum: | |
|---|---|
| **Avatar (L1):** | An avatar is a 3-dimensional character that represents a physical person in social MR environments. While it may not be an exact physical replica, the avatar typically resembles the person in terms of appearance (morphological similarity) and movement (kinematic similarity), creating a digital representation of the individual. |
| **Semi-Autonomous Virtual Character (L2):** | A Semi-Autonomous Virtual Character is a hybrid between an Avatar (L1) and an Autonomous Virtual Character (L3). It is termed semi-autonomous because its movements and behaviors are partially controlled by a real person, with the remaining actions generated by algorithms, such as artificial intelligence or automated rules. |
| **Autonomous Virtual Character (L3):** | An Autonomous Virtual Character is an embodied autonomous agent. An embodied agent is an agent that interacts with other entities in a social MR through a physical body within that space. An autonomous agent is a system situated within, and as a part of, a social MR that senses that space and acts on it, over time, in pursuit of its own goals. |
| **Autonomisation Continuum:** | These three digital agents correspond to three successive levels of autonomisation (autonomy in artificial agents): avatars (L1) are directly driven by (usually remote) human agents and are simply reproducing their movements with a certain degree of realism; Semi-autonomous virtual characters (L2) are driven by human agents, but can adapt, for instance through movement amplification or attenuation, parts of their behaviour; Autonomous virtual characters (L3) have the full degree of autonomy through their AI-powered cognitive architecture. |

| Theme 2 - Technical challenges | |
|---|---|
| **Ego-centric visual-inertial tracking:** | A technique used to estimate the position and orientation of human body segments via a combination of visual and inertial sensors. In this approach, the camera is mounted on the head-mounted display (ego-centric perspective) and a minimal number of additional sensors are placed on the self-occluded segments (lower-body/legs). |
| **Scene neural rendering:** | An emerging class of image and video generation approaches based on deep learning that enable synthesizing images from real-world observations. It leverages generative machine learning, Neural Radiance Fields (NeRF), or Gaussian Splatting techniques and allow creating high-fidelity photo-realistic images and videos of complex scenes. |
| **Movement representation:** | Unsupervised or semi-supervised machine learning approaches for dimensionality reduction of sequential input data, such as high dimensional movement kinematics, allowing to understand and reconstruct/create or appropriate animations for avatars/virtual (semi-) autonomous agents in social MR. |

| Theme 3 - Scientific challenges | |
|---|---|
| **Social connectedness:** | The psychological feeling of inclusion or acceptance into a group of human and virtual agents interacting together in a group. |
| **Synchronisation:** | A condition of alignment regarding the motion of individuals and/or virtual humans, expressed as time signals. Motions are synchronized if all signals are equal in time ("phase-locked"), except possibly for a small difference and/or a constant time delay. |
| **Sensorimotor primitives:** | The building component of bodily actions by intentional agents, consisted of coordinated kinematic variables, dynamic variables, and sensory variables. |
| **Sensorimotor propagation:** | The transmission and entrainment of social information, coded in sensorimotor primitives, across avatars and/or virtual (semi-)autonomous agents. The degree of propagation is rooted into the amount of social information encoded and transmitted. |
| **Amplification - attenuation of motor primitives:** | Amplification of sensorimotor primitives allows to encode social information in such a way that it facilitates the readout of social information by human interactants. The process of attenuation allows to dampen social information transmission by not encoding social information in the sensorimotor primitives. |
| **Embodiment:** | The pre-reflective experience combining sense of self-location in the MR space (i.e., I am located where my avatar is located), sense of agency (i.e., I am in control of the actions of my avatar) and sense of ownership (i.e., my avatar body is my body). |
| **Social presence:** | The sense that the experiences rendered in MR are authentic (other virtual humans collocated in the same environment as the user are volitional) and that users feel connected to their virtual representation (within layers of body, emotion, and identity). Social presence encompasses both psychological and physical sense of 'being' in the shared hybrid space anchored in a current moment. |

| Theme 4 - Ethical challenges | |
|---|---|
| **L1 Sociality:** | The social, cultural, biological and personal bases to human coexistence as a social animal relevant to reproduction and survival. A digital replica acts as a proxy for the user and subsequently is an extended vehicle for sociality. |
| **L1 Dignity:** | The notion of dignity encompasses the idea of basic worth afforded to individuals and protected by law. Digital replicas are extensions of the personhood(s) of the users and therefore are constrained and influenced by social norms and practices. |
| **L2 Self-Identity:** | The ontological state of self-awareness, which differentiates one individual from another and marks out self-ownership. The sense of self has to be preserved and/or protected when humans use semi-autonomous modes. |
| **L2 Transparency:** | In Social MR, transparency will concern awareness and knowledge about the cognitive architecture used for both human and autonomous agents and the moderation of their interactive dynamics. |
| **L2/L3 Responsibility:** | Offsetting future harms by conscious and ethical deliberation and implementing cautionary approaches such as "do not do significant harm". In Social MR, this can mean considering potential or undesirable outcomes arising from its expansion to broader applications. |
| **L2/L3 Cooperation:** | Cooperation is a fundamental aspect of human sociality, understood as reciprocal acts shaped over time for mutual benefit. In Social MR, it will be the key alignment principle of human and autonomous characters' interactions. |
| **L3 Attachment:** | Decreasing face-to-face interactions in favour of computer-mediated interactions means social entanglement with virtual autonomous characters. This might hold implications for attachment characteristics such as human social skills, emotional regulation, and bonding |
| **L3 Trust:** | There is an inherent opaqueness of AI systems, leading to potential issues of deception and humans being unable to predict what autonomous agents will do, and how reliable and secure the given system is. To promote synthetic cooperation between humans and autonomous characters, it may be vital to know the identity of agents, their capabilities and their interaction roles. |

## 6.   <u>REFERENCES</u>